\documentclass{article}
\usepackage{amsmath} 
\usepackage{amssymb}
\usepackage{epsfig} \usepackage{dcolumn} \usepackage{array} \usepackage{verbatim}  

\newcommand{\cM}{\mathcal{M}}

\newcommand{\st}{\textit{star}}
\newcommand{\li}{\textit{link}}
\newcommand{\gn}{\textit{gn}}

\newtheorem{theorem}{Theorem}
\newtheorem{lemma}{Lemma}
\newtheorem{corollary}{Corollary}

\newcommand{\proof}{\paragraph{Proof:}}

\newcommand{\qa}{(\textit{a\/}) }
\newcommand{\qb}{(\textit{b\/}) }
\newcommand{\qc}{(\textit{c\/}) }
\newcommand{\qd}{(\textit{d\/}) }

 \newcommand{\dc}{C_d}  
 \newcommand{\cc}{C_c}  

\newcommand{\TA}{\textit{Tri}_1}
\newcommand{\TB}{\textit{Tri}_2}
\newcommand{\TC}{\textit{Tri}_3}

\newcommand{\om}{\omega}

\newcommand{\eg}{\textit{e.g.\/}}
\newcommand{\ie}{\textit{i.e.\/}}

\font\openface=msbm10 at10pt

\def\Reals         {{\hbox{\openface R}}}

\begin{document}
\title {Coarse-graining dynamical triangulations: \\ a new scheme}

\author{Joe
Henson\footnote{Perimeter Institute for Theoretical Physics,
31 Caroline St. N, N2L 2Y5, Waterloo ON, Canada}}
\maketitle

\begin{abstract}

A new procedure for coarse-graining dynamical triangulations is
presented. The procedure provides a meaning for the relevant value
of observables when ``probing at large scales'', \eg{} the average
scalar curvature.  The scheme may also be useful as a starting
point for a new type of renormalisation procedure, suitable for
dynamically triangulated quantum gravity.

Random Delaunay triangulations have previously been used to
produce discretisations of continuous Euclidean manifolds, and the
coarse-graining scheme is an extension of this idea, using random
simplicial complexes produced from a dynamical triangulation. In
order for a coarse-graining process to be useful, it should
preserve the properties of the original dynamical triangulation
that are relevant when probing at large scales. Some general discussion of this point is given, along with some arguments in favour of the proposed scheme.

\end{abstract}

 \vskip 1cm

\section{Introduction}

In the candidate theory of quantum gravity (QG) known as causal
dynamical triangulations (CDTs)
\cite{Ambjorn:2005qt,Ambjorn:2005jj,Ambjorn:2006jf,Ambjorn:2008zza}, a regulator
is introduced to ``cut out'' short distance detail from the model
and render the physical quantities finite.  To obtain a continuum
theory the regulator has to be sent to zero.  At the same time,
some kind of renormalisation is expected to be necessary in order
to recover finite results for physical quantities.  Without
renormalisation, for example, the average scalar curvature in the
4-dimensional CDT model diverges.

Renormalisation depends on being able to identify actions (or
Hamiltonians) that represent the same physics at large scales, but
with different cut-off scales.  This can be done by identifying,
and then integrating over, degrees of freedom that are irrelevant
when probing the system at large scales.  For example, when
probing water with optical light, density fluctuations near the
atomic scale are irrelevant in this sense.

In order to do this, a coarse-graining scheme is necessary: from a
``fine-grained'' history with a certain cut-off, a coarse-graining
is found with a lower cut-off, but with corresponding properties
at large scales.  For lattice quantum field theory (QFT), one way
to achieve this is by the Kadanoff block spin transformation (see
\eg{} \cite{Wilson:1974mb,Swendsen:1979gn}).  By such techniques,
the high frequency modes of a field can be removed from a history
without significantly affecting low frequency modes.  This accords
with the idea of removing small scale fluctuations that are not
relevant when probing on large scales.

In view of this success, it would be of interest to extend these
ideas to CDT quantum gravity.  Previous attempts have been made to
introduce coarse-graining to Euclidian dynamical triangulations
\cite{Renken:1994nq,Renken:1994gi,Renken:1996kf,Renken:1996ga,Thorleifsson:1995ki,Ambjorn:1996hu}.
Also, in the spin-foam quantum gravity program
\cite{Oriti:2001qu,Perez:2004hj}, some preliminary studies of the
problem have been made, discussing differences between quantum
field theory coarse-graining and quantum gravity coarse-graining,
and suggesting the use of some interesting algebraic structures to
extend the standard renormalisation group techniques
\cite{Markopoulou:2002ja,Oeckl:2002ia,Oeckl:2004yf}. In the causal
set approach \cite{Bombelli:1987aa,Henson:2006kf} there is a
natural coarse-graining procedure that transforms the discreteness
scale \cite{Sorkin:1990bh,Rideout:2000fh}; some of the statistical
techniques used below are similar to those employed in the causal
set case.

For CDTs, the most immediate use of coarse-graining ideas would be
the definition and calculation of coarse-grained observables, for
comparison to simple observations and accepted theory.  The theory is uniquely advanced in this regard: the expectation values of some interesting observables have already been computed, with good results (see \textit{e.g.} \cite{Ambjorn:2005db,Ambjorn:2008wc}).  However, it has so far been difficult to find a large number of physically relevant observables to compute.

In closest analogy to the QFT case, in quantum gravity theories
the idea would presumably be to take our regularised (or
fundamentally discrete) histories, and produce from each one a
coarse-grained version.  This idea of coarse-graining each history has been very useful until now in fixed-background coarse-graining, and none of the generalisations required by quantum gravity suggest any new reason to abandon it (see \cite{Henson:2009yb}, where this point is discussed).  As suggested by the brief comments above,
it is absolutely crucial that, on large scales, this coarse-graining
corresponds to the original fine-grained version on large scales.
This must be the first concern of any such scheme, as without this
property the coarse-graining is of no physical significance.

This criterion of ``physical aptness'' is more problematic in quantum gravity than in standard cases.  In simple cases of lattice field theory, the problem is so close to trivial that it is barely discussed.  There is a straightforward prescription that tells us what features of a history are relevant
when probing at large scales (they are simply the low-frequency modes of the field), and many different block spin
coarse-graining schemes can be found which adequately preserve these features. In lattice QCD things are less trivial; some coarse-graining schemes better encode the degrees of freedom that are relevant at larger scales.  In quantum gravity, the notions of large-scale and small-scale themselves become more subtle due to the dynamical nature of spacetime. Because of this it becomes necessary to consider the question ``when do
two geometries correspond at large scales?'', and provide a reasonable physical justification for the answer.  For example, a
candidate answer to this question, based on properties of the
Lapacian operator, is a cornerstone of the successful application
of renormalisation group ideas to Euclidean quantum gravity
\cite{Reuter:1996cp,Lauscher:2005xz,Percacci:2007sz}. This approach has produced
evidence for the non-peturbative renormalisability of the theory
in 4D.

When designing a coarse-graining scheme for quantum gravity, a number of new requirements arise, some stemming from the non-trivial nature of this physical aptness requirement.  These present difficulties for any na\"ive blocking procedure, as discussed below in section \ref{s::coarse}.  On the basis of these considerations, a new coarse-graining scheme for dynamical triangulations, suitable for use in the CDT program, is presented here. The procedure improves on previous ideas, as it arguably accords better with the physical criteria discussed below.  From this, large-scale effective versions of many observables, such as the average scalar curvature, can be defined.  The class of observables also provides the basis for a scheme that can be used to define effective actions for CDT theory, at least in principle.  Most of this discussion is complementary to other recent studies, which were largely concerned with the general mathematical tools necessary for coarse-graining in quantum gravity.

As the proposed scheme has many novel aspects, a full explanation is made here without also including computational results, which will be left for future work. In section \ref{s::coarse}, the new problems of coarse-graining for a theory of dynamical geometry are discussed.  The coarse-graining scheme is then presented in section \ref{s::newscheme}, along with some coarse-grained observables of particular interest.   In section \ref{s::practical}, a simplified version of the coarse-graining process that is practical to apply to CDT simulations is given, and there is some discussion on how to apply it.

\section{Coarse-graining and dynamical geometry}
\label{s::coarse}

As already noted, physically relevant observables are hard to come by in lattice QG, as there are a number of difficult criteria that they must satisfy in order to be useful. These will be of relevance throughout the
paper. The observables must be:

\begin{list}{}{}
\item[\qa] well defined generally covariant observables for the fundamental
theory;
\item[\qb] possible to define in the discrete setting of
CDTs;
\item[\qc] practical to calculate (so far this has meant by
computer simulation);
\item[\qd] relevant when probing the system at large scales.
\end{list}

Clearly, $\qd$ entails that the observables do not diverge to
infinity in the continuum limit. Perhaps the best existing example
is the spectral dimension \cite{Ambjorn:2005db}, which is a measure of the effective
dimension derived from the spectrum of the Lapacian operator.  This will serve as an example of the kind of observable we are searching for.   However, the spectral dimension was not derived from a general scheme \footnote{However, it is true that much more information is encoded in the spectrum of the Lapacian than just the dimension.  This may provide a different way to define interesting observables beyond the spectral dimension, and to compare results of CDT simulations to simple classical geometries. These questions are currently under investigation by the author and Dario Benedetti.}.

Before attempting to define our observables and to justify them against the
conditions set out above, it is necessary to consider the physical
requirements, especially $\qd$.

\subsection{Geometry and Scale}
\label{s::geometry}

In the final analysis, the meaning of ``properties relevant at large/small scales'' could be determined by detailed physical arguments, for example involving gadanken-experiments.  However, usually a more practical approach is taken.  In the fixed background case, the distinction is made by considering field modes of different momenta.  Block-spin transformations do not exactly preserve the amplitudes of these field modes, but they have been proven to be successful by producing non-trivial, plausible results, and then by comparison to each other and to experimental data.  Thus the physical aptness of a coarse-graining is not usually derived from first principles, but by a process of physical insight combined with experimental (and computational) input.  However, the guiding principle that large scales are associated with small field momenta plays a crucial role.

In the QG case we are just beginning to retrace these steps.  Without a fixed background geometry we must generalise even the simple guiding principle mentioned above, which was previously the easiest part of the argument.  In this vein, it would be useful to have at least some conception of the following:  a reasonable measure of the scale on which two geometries differ.  The measure could be based on spectral properties as in the recent application of the renormalisation group to gravity, or perhaps applying the Gromov-Hausdorff distance \cite{Gromov:1981,Ambjorn:1996ny} between two geometries.  This would raise some difficult questions.  For example, how should the kind of ``disordered locality'' discussed in  \cite{Markopoulou:2007ha} be treated?  Also, should the measure of closeness (and hence the coarse-graining cut-off) be frame dependent, or should it respect general covariance?  Usually it is taken to be frame dependent, to the extent that a foliation of spacetime is used to perform the Wick rotation, and coarse-graining proceeds on the Euclidean side.  A truly covariant coarse-graining, on the other hand, would presumably mean that small \textit{volume} behaviour is cut-out, hopefully leading to high momentum behaviour being suppressed in the frame relevant to any interaction (the causal set coarse-graining, which depends only on causal and volume information, is a possible example of this \cite{Sorkin:1990bh}).  Below, the standard route used for fixed-background cases is followed, \textit{i.e.} to consider issues of scale after Wick-rotation, but apart from that to preserve covariance as much as possible.

Even if the coarse-graining scheme itself is not covariant on the Lorentzian side (as in standard cases), care must be taken to apply it in a way that respects the symmetries of the theory -- the relevant scale of coarse-graining should be set at a physical scale relevant to the system under consideration, \textit{e.g.} centre-of-mass energy, and therefore must never be decided by reference to any non-dynamical frame or foliation.  Then the calculated observables should be approximately covariant.  This is how condition \qa of section \ref{s::coarse} is met in practice.  While there is no guarantee that this approach can be extended to QG, it is a useful working hypothesis.  The foliation chosen in CDT simulations is thought to roughly correspond to foliation by cosmological time.

At this early stage, a detailed discussion of this measure of closeness of geometries can be deferred.  Finding one coarse-graining scheme that gave good results for CDTs would be an advance, and could help to guide further progress in defining the conditions themselves.  But these considerations do lead, at least, to some necessary conditions for a QG coarse-graining scheme to be physically apt.  As part the general problem of separating large and small scales properties, some new pitfalls arise in this case.  These are discussed in the following subsection.

\subsection{Some pitfalls for QG coarse-graining}
\label{s::pitfalls}

There are some key differences between the QG and standard cases of coarse-graining, some of which have been noted in previous studies.  Firstly, a coarse-graining for discrete QG cannot preserve the lattice structure.  Secondly, quantum geometry can give radical departures from the classical geometry that we are used to.  Dimension, as shown in  \cite{Ambjorn:2005db,Ambjorn:2005qt}, is now a scaling property in the CDT model.  The same would probably be true of topology, if dynamical topology was allowed.  Indeed, at some scales it is not clear that the geometry should closely resemble an extended smooth spacetime at all, to any extent beyond the fact that it is also a metric space.  If the coarse-graining scheme does not reflect this, we may already have fallen foul of a classical prejudice.

Thirdly, the coarse-graining scheme should not preserve the average scalar curvature (at least in more than 2 dimensions\footnote{In 2 dimensions
the situation is rather different, as a consequence of the
Gauss-Bonnet theorem. This may be why preservation of average
curvature is looked upon more favourably in
\cite{Thorleifsson:1995ki}, where a coarse-graining scheme for 2D
dynamical triangulations is developed.}).  Most importantly, this is because this quantity has been seen to diverge in the case of CDTs.  Also, this does not seem consistent with a coarse-graining that preserves large scale features of the geometry.  It is not difficult to come up with examples of two geometries that differ only in many small volume, non-overlapping regions (or are otherwise intuitively ``close''), but have arbitrarily different average scalar curvature.  As well as the divergence, this speaks against preserving average curvature under coarse-graining.

Forth, there is a problem that applies particularly to DTs.  By decreasing the lattice spacing, any Euclidean geometry can be arbitrarily well approximated by a dynamical triangulation (a precise theorem to this effect can be found in \cite{Ambjorn:1996ny}).  Because of this, they are a useful way to discretise the space of geometries for quantum gravity.  However, there is no equilateral triangulation of flat space above 2D, only dynamical triangulations that approach it as the lattice spacing $a \rightarrow 0$.  The same is true of any space that has low curvature on large scales -- exactly the type of geometry we hope to obtain from coarse-graining.  This might not seem problematic; after all, in all lattice approximations, the degree of approximation to continuum configurations depends on the lattice length.  But it does compromise a coarse-graining scheme that merely increases the lattice length for dynamical triangulations.  In this background independent case it is highly non-trivial to identify those DTs that are close to \textit{e.g.} flat space (for example, the value of the Regge action of a DT that is close to some low-curvature geometry may not be correspondingly close the the value of the Einstein-Hilbert action for that geometry).  Without a means of doing so the usefulness of the coarse-graining scheme is in doubt, since it cannot produce observables that are relevant at large scales.

Any good coarse-graining scheme for CDTs needs to avoid these problems, and most of the problems are also relevant for other approaches such as spin-foams.  Previously considered coarse-graining schemes were very similar to Kadanoff coarse-graining.  The general idea was to replace a block of many simplices with a block containing fewer simplices, which were the same except for the lattice scale.  This avoids the first pitfall, but not the others.  This kind of blocking does not allow the dimension to vary, or allow the coarse-graining be anything other than a triangulation of a manifold.  Some schemes advocated preserving average curvature.  Also, schemes for dynamical triangulations have always produced dynamical triangulations as output.

Some of these problems might be overcome by a more sophisticated treatment of the continuum approximation, if it were possible to identify smooth approximations to DTs in some systematic way.  However, even if this could be achieved, this is not a very natural approach.  Geometries of many different dimensions would be represented by DTs with some particular fixed dimension, for example.  The new coarse-graining process presented below is designed with these issues in mind.

\section{The new scheme}
\label{s::newscheme}
\subsection{Delaunay complexes}
\label{s::delaunay}

How can we extract such large-scale information from a given
geometry? An answer to this question is a step towards a
physically apt coarse-graining scheme. Fortunately, some similar
ideas have been developed in other contexts, where statistical and
combinatorial techniques have been used to make discrete
approximations to Euclidean geometries, with controllable cut-off
scales.  The relevant concept is the Voronoi procedure, which
produces discrete structures from a metric space. The large scale
properties of a continuous Euclidean signature space are
conjectured to be encoded in these so-called ``random Delaunay
triangulations'' of that space \cite{Bombelli:2004si}. This
procedure has previously been applied to the problem of continuum
approximations for spin network configurations in loop quantum
gravity \cite{Bombelli:2004si}.  Outside quantum gravity, similar
techniques are important for the random lattice formulation of QFT
\cite{Christ:1982zq,Lee:1983ns}. Before the application is given,
the definition of the procedure will be stated and some comments
made as to how the resulting structures encode large-scale
geometry.

First, a note on terminology is necessary.  In general, the
terminology is standard to dynamical triangulations (see \eg{}
\cite{Ambjorn:1996ny}), with the proviso that simplicial complexes
are here considered in the abstract definition only. An abstract
simplicial complex $C=\{S_0,S_1,...,S_{D-1}\}$ is a set of
vertices, or ``0-simplices'', $S_0$ and non-empty sets $S_n$ of
$(n+1)$-tuples of vertices, or ``$n$-simplices''.  $D$ is the
dimension of the simplicial complex.  To be an abstract simplicial
complex, all faces of simplices (\ie{} subsets of simplices) must also be simplices.
As the alternative geometrical description of simplicial complexes
will not be used in this paper, the prefix ``abstract'' is dropped
below.  The $n$-skeleton of a simplicial complex $K$ is the
sub-complex $K_n \subset K$ made up of all simplices in $K$ of
dimension $\leq n$. For example, the $1$-skeleton is made up of
edges and vertices only, and as such is a graph. Where no
confusion arises, the 1-skeleton will be referred to as the
skeleton.

Consider a Euclidean geometry $\cM$ and a collection of $N_c$
points $P$ in $\cM$. A simplicial complex called the
``Delaunay complex'' $\dc(P,\cM)$ can be associated to the
points. The points are the vertices in the complex, and simplices
are made up of sets of nearby vertices, according to the following
prescription.

Associated to each point $p \in P$ is an open region $\om(p)\subset
\cM $, its ``Voronoi neighbourhood'', made up of all points in
$\cM$ that are closer to $p$ than any other point in $P$.
In the Delaunay complex, an edge is placed between two vertices
$p$ and $q$ iff the closures of $\om(p)$ and $\om(q)$
share any points (\ie{} if these regions border each other).  For
higher dimensional simplices, the prescription is similar: an
$n$-simplex is placed between a set $v \subset P$ of $n+1$ vertices iff

\begin{equation}
 \bigcap_{p \in v} \overline{\om(p)} \neq \emptyset,
\end{equation}

where $\overline{\om(p)}$ signifies the closure of $\om(p)$.  If
the points in $P$ are selected uniformly at random according to
the volume measure on $\cM$, we call the resulting simplicial complex a random Delaunay complex $\dc(N_c,\cM)$.

If $\cM$ is flat or uniformly curved, then in this random process,
it is with probability 0 that any point in $\cM$ is equidistant
from more than $D+1$ points in $P$, where $D$ is the
dimension of $\cM$ \cite{Bombelli:2004si}.  For any manifold, if
the points are ``at generic locations and sufficiently dense''
\cite{Bombelli:2004si}, the resulting complex $\dc(P,\cM)$
is a triangulation of $\cM$.  This means that, if the
``sprinkling density'' $\rho$ ($N_c$ divided by the volume of
$\cM$) is sufficiently large with respect to the maximum
curvature of $\cM$, then the random Delaunay complex
$\dc(N_c,\cM)$ will be a triangulation with high probability.
Below we refer to such manifolds as ``low curvature on the
sprinkling density scale''.

For higher curvatures this may not be true. As a visually
accessible example, consider a 2-sphere connected to another
2-sphere by a thin neck, such that the volume of the neck is much
less than that of the spheres. Below a certain density of
sprinkled points, it is unlikely that the neck will contain any
sprinkled points. Instead, it will probably be covered by two
Voronoi neighbourhoods, one associated to a sprinkled point in
each sphere. Thus, there will be an edge between these points.
This will be the only edge connecting two points sprinkled into
different spheres.  The Delaunay complex will therefore contain an
edge that is not part of any triangle. This is the reason that
$\dc(P,\cM)$ is referred to here as a Delaunay
\textit{complex} here, rather than the more commonly used
``Delaunay triangulation''.

We are interested in the random Delaunay process because the
complexes $\dc(P,\cM)$ produced are arguably insensitive to
small scale detail of $\cM$.  However, there are some
counter-examples to this involving non-trivial topology (\textit{e.g.} holes).  The application to the presently considered CDT
models will not be affected by this, and so the problem is not addressed in detail here.  For now attention will be restricted to cases without complicated dynamical topology (although one can consider modifying the procedure to cope with microscopic topological detail, a subject that will be dealt with elsewhere).

\subsection{Delaunay Complex Observables}

\label{s::observables}

A large class of observables of a Euclidean Geometry $\cM$ can be
defined using the idea of the Delaunay complex of finite sets of
points in $\cM$ (we will assume $\cM$ to be of finite volume
$V(\cM)$ here). First, consider any real-valued function of
(isomorphism classes of) simplicial complexes, say $f(S)$.  This
function holds a value for the Delaunay complex
$\dc(P,\cM)$, but this value depends on the positions of the
points $P$. To form an observable $O_f$, we must average
over all positions of the points:

\begin{equation} \label{e::observables} O_f=\int_\cM d^Dp_1
\sqrt{g(p_1)} \int_\cM d^Dp_2 \sqrt{g(p_2)} ... \int_\cM
d^Dp_{N_c} \sqrt{g(p_{N_c})} f \bigl( \,\dc(P,\cM)\,\bigr),
\end{equation} where $D$ is the dimension of $\cM$ and $P=\{p_1,p_2,...,p_{N_c}\}$.

These are dubbed ``Delaunay complex observables''.  This kind of
observable would clearly be extremely difficult to calculate
analytically, but may be possible to approximate.  The observable
$O_f$ is the average value of $f \bigl( \,\dc(N_c,\cM)\,\bigr)$
for random Delaunay complexes $\dc(N_c,\cM)$, and so the random
Delaunay complex procedure could in principle be used to sample
the value of $f \bigl( \,\dc(P,\cM)\,\bigr)$, as a means to
estimate $O_f$. If the observable is relevant in the classical regime, it should have low variance, and so this random sampling would quickly converge to the correct value for $O_f$. Indeed, there is already
evidence that a single sample may be enough to accurately
calculate some observables \cite{Bombelli:2004si}.

As manifestly generally covariant quantities, the observables
satisfy criterion \qa given at the beginning to section \ref{s::coarse} above (although after Wick rotation; see section \ref{s::geometry}).  Criteria \qb and \qc will be dealt with later by giving a discrete version of the
Delaunay process that is suitable for CDT computer simulations.

The conjecture is now made that a large class of Delaunay complex
observables satisfy criterion $\qd$, as discussed in section
\ref{s::geometry}.

There is evidence from previous studies that large-scale geometrical information is captured by such observables.  In previous work on the random Delaunay complex
procedure in curved spaces, Bombelli, Corichi and Winkler consider
manifolds that are low curvature on the sprinkling density scale.
They conjecture that, from random Delaunay complexes
$\dc(N_c,\cM)$, the geometry of the original manifold $\cM$ can be
approximately reconstructed \cite{Bombelli:2004si}. This claim is
given justification in 2D.  For example, they have shown that the
average curvature of a 2D geometry can be calculated from the
expected average valency of the vertices of a random Delaunay
complex on that geometry, and is found to obey

\begin{equation}
 \label{e::2dcurvature} R(S)=4 \pi \rho ( 1 - \frac{1}{6} \bar{N_1} )
\end{equation}
where $R$ is the average curvature of a geometry $\cM$, $\rho$ is
the sprinkling density $N_c/V(\cM)$, and $\bar{N_1}$ is the mean
valency of vertices in a randomly generated complex
$\dc(N_c,\cM)$. Note that this curvature estimator is only a
function of the skeleton of the complex.  Similar results exist in 3D \cite{isokawa:2000}, for negative curvature.  It is conjectured that the valency of vertices will also be a function of the scalar curvature in higher dimensions, for manifolds with small curvature
on the sprinkling density scale as discussed above\footnote{There
is now good computational evidence for this conjecture as applied to spheres
in 3 and 4 dimensions, which will be presented in future work.}.

This speaks for the preservation of large scale information when
the fine-grained geometry is low curvature.  Since the Delaunay
complexes are discrete, each one contains only a small fraction
of the information of the fine-grained geometry, and so it is
reasonable to conjecture that small-scale detail is absent from
the coarse-graining, as required.

As in all such schemes, the final test will be in the calculation.  It may not be true that a small deformation (appropriately defined) does not affect the placement of edges in the Delaunay complex; for example, faces of microscopic size might appear due to exotic geometry near the Planck regime.   Hopefully, the question of whether this happens in CDT theories can be settled by simulations, and the coarse-graining procedure could be refined if necessary.

At large values of $N_c$, a single coarse-grained simplicial complex is conjectured to be enough to approximately reconstruct the geometry of the mainfold it was generated from.  Such a complex can be given a geometrical interpretation of its own \cite{Bombelli:2004si}.  The interpretation of such a complex is different from the DT interpretation, which attributes equilateral geometry to all simplices.  Here the simplicial complex corresponds to a continuum geometry if and only if it could have come from the Random Delaunay process on that geometry with ``relatively high probability'' .  This rule does not uniquely fix the geometry without further refinement, but this is not crucial.  For example, for one random Delaunay complex, the local scalar curvature at a marked point might vary over all corresponding geometries allowed by his definition, making it badly approximated, but this is not a quantity of observational interest (it has no relevance at large scales).  The effective average curvature of a small region will be well approximated.

As mentioned above, dynamical triangulations with large lattice length do not approximate well to smooth spacetimes.  It may seem odd, therefore, to use a similar structure (an unlabelled simplicial complex) for the coarse-grainings.  However, using the alternative geometrical interpretation given here, which is natural for random Delaunay complexes, avoids the problem with DTs.  With this approach, low curvature spacetimes can be approximated, and it is possible to read off observables of relevance at large-scales from the discrete structure.  This is not the case for DTs; no criterion exists to check if a DT is close to a flat space, for instance.  It is important to note that, even when the coarse-graining is a triangulation (which is not always the case), this geometrical interpretation may differ from the interpretation as a dynamical triangulation.  There is no obvious reason for the two interpretations of unlabelled triangulations to coincide for all observables (although it is true the expression for average curvature is the same in 2D).

\subsection{Some useful coarse-grained observables}
\label{s::examples}

Of the large class of observables defined above, some are of particular interest for the study of CDT quantum gravity simulations.  The \textit{effective average scalar curvature} is an example.  Assuming the main conjecture of the previous section is correct, we
may call the curvature estimator $R(S)$ above, or rather its Delaunay complex observable $O_R$, the effective average scalar curvature.  Also, we can consider definitions of \textit{effective dimension}. These random Delaunay complexes have a dimension, which provides new effective dimension estimators as Delaunay complex observables.  For each vertex in a complex there is a maximum dimension of simplex which contains it. There is also a minimum dimension of simplex that contains the
vertex but is contained in no other simplex.  The average value of
these numbers over all vertices provide two fractal dimension
estimators, giving alternatives to the spectral and Hausdorff dimensions.
Hopefully the new dimension estimators will agree with the other
ones for effectively manifoldlike geometries (although for other
geometries they may differ without this being problematic).

As well as familiar observables like scalar curvature, some novel observables are of interest in quantum gravity.  Spacetime can be so
curved that, on some scales, there may be no smooth Euclidean manifold at
all which would qualify as a good coarse-graining under sensible
rules. In this case, more general coarse-grained observables than the above may be useful to judge whether the simulations are approaching a ``manifoldlike'' regime at large scales.  We need effective measures of \textit{manifoldlikeness}.

Consider a geometry $Y$ that is low curvature on a certain sprinkling density
scale. It has been recalled that a random Delaunay complex on this
geometry is with high probability a triangulation with the
dimensionality of $Y$.  According to the conjecture, a manifold $X$
that approximates $Y$ on large-scales would have this
property also.  Thus, the property of a random Delaunay complex
being a triangulation is a measure of manifoldlikeness.  It would be interesting to ask how close to a triangulation the Delaunay complexes are for a particular geometry, as a better measure of effective manifoldlikeness at large scales. This
requires a measure of how like a 4D triangulation a complex is.
Some observables of this type are given for 2D in appendix
\ref{a::2Dtriangulations} which suggest generalisations to higher
dimensions.

\subsection{Effective actions}

\label{s::coarseaction}

The main aim here is to define some observables that can be used to probe the large-scale features of the CDT model.   We can also consider using the coarse-graining scheme to define an effective action.  The procedure, and purpose, of defining an effective action in a fixed-background lattice theory is well known.  A treatment in a similar context to the present one is given in \cite{Renken:1994nq}.  Basically, the partition function on the model is rewritten as a sum of contributions from coarse-grained configurations\footnote{In the above scheme, a probability distribution over simplicial complexes is defined from each fine-grained configuration.  This may be a rather unweildy object to consider as a coarse-graining.  When $N'$ is large, however, observables like curvature can (with high probability) be accurately estimated by taking
only one sample from the random Delaunay complex process, and so one sample can be considered as ``the coarse-grained configuration''.}, which are in turn calculated by summing contributions from the original fine-grained configurations.  The CDT theory can be written in terms of a statistical sum over geometrical
configurations $T$ weighted by $e^{- S(T)}$ where $S$ is our
Euclideanised action.  At fixed volume, the partition function may be written briefly (see \textit{e.g.}  \cite{Ambjorn:2005qt}) as
\begin{equation}
 \label{e::CDTpi}
 Z_a(V,G)= \sum_T \mu(T) e^{- S(T)},
\end{equation}
where $Z_a(V,G)$ is the partition function which depends on $V$ the
total volume, $G$, the bare Newton's constant, and the lattice
spacing $a$. The action $S(T)$ and measure factor $\mu(T)$ are
defined elsewhere \cite{Ambjorn:2005qt}.  The sum is over all
causal dynamical triangulations $T$ in a certain class (for instance
4D CDTs with $S^3 \times \Reals$ topology).  The effective action takes the form

\begin{equation}
\label{e::effectiveS2}
e^{- S_{N_c}(C)} = \frac{1}{ \mu(C)} \sum_T Pr(C, T) \mu(T) e^{- S(T)},
\end{equation}
where the sum is as above, and $Pr(C, T)$ is the probability of
generating the simplicial complex $C$ from the random Delaunay
complex procedure on these geometries:

\begin{equation}
Pr(C, T)=
\int_\cM d^Dp_1
\sqrt{g(p_1)} \int_\cM d^Dp_2 \sqrt{g(p_2)} ... \int_\cM
d^Dp_{N_c} \sqrt{g(p_{N_c})} \,
\delta \bigl(C, \,\dc(P,\cM) \, \bigr),
\end{equation}
where $D$ is the dimension of $\cM$, $P=\{p_1,p_2,...,p_{N_c}\}$, and $\delta(C_1,C_2)=1$ if graph $C_1$ is isomorphic to $C_2$,
and $0$ otherwise (note that $P(C, T)$ is a diffeomorphism
invariant quantity for these Euclidean geometries).  The random Delaunay complex procedure is defined for any metric space with volume measure, and so, in principle, it could be applied to DTs directly (although a more practical scheme is given in section \ref{s::practical}).  This is an effective action on the space of all simplicial complexes.

In this scheme, the scale is controlled by the sprinkling density
$\rho=N_c/V$, where $V$ is the volume of the fine-grained
geometry.  Rescaling after coarse-graining would be taken care of by using appropriately scaled values of $\rho$.

Note that here, due to the irregular and dynamical lattice, it is not obvious how to break  $Pr(T', T)$ down into a product of terms, one for each ``block'' (whatever that would mean), as is possible in the standard case (indeed, this feature makes it easier to find schemes that preserve large-scale properties there).  Although this is a disadvantage, it is not clear if any scheme could avoid it while still meeting the criteria for a good coarse-graining scheme, as discussed above.

Now the coarse-grained histories are general simplicial complexes,
to be interpreted as the results of the random Delaunay complex
procedure. The fine-grained histories on the other hand are
DTs.  Iterating the scheme as in the fixed background case is therefore problematic.  This does not destroy the usefulness of the scheme, however.  The coarse-graining can still be applied for different values of $\rho$ to obtain coarse-grainings at different scales
(similarly to the techniques used to calculate the spectral
dimension).  The values of observables could still be checked at
different scales to search for a fixed point and/or obtain their
physical values in the continuum limit, and this is the main goal
of this study.

\section{A practical scheme}

\label{s::practical}

In previous sections, a way to coarse-grain dynamical
triangulations has been given. It was commented that the Delaunay procedure could be applied to DTs considered as piecewise-linear manifolds.  However, it is likely to be impractical to implement by computer, since it would be necessary
to calculate the geometrical distances between general points in a
dynamical triangulation.  Because of this, it is useful to
consider an approximate, discrete version of the procedure that
can be applied to a dynamical triangulation,
satisfying criteria \qb and \qc of section \ref{s::coarse}.

These modifications of the scheme should be compared to using random walks on the simplices to calculate the spectral dimension, rather than a continuous diffusion process \cite{Ambjorn:2005db}.  It is a reasonable conjecture that the simplifying modifications
of the simplicial coarse-graining procedure will not significantly
effect results, particularly when the coarse-graining ratio
$N_f/N_c$ is large.

\subsection{Coarse-graining a simplicial complex}

\label{s::coarseg}

A procedure for coarse-graining a simplicial complex is now
presented.  The algorithm is specified to the extent that concrete implementation by computer is possible.  The procedure is modelled on the random Delaunay complex process, except that it uses the combinatorial properties of the
triangulation and avoids use of the geometrical distance.

The fine-grained history is taken to be a simplicial complex or
dynamical triangulation\footnote{In some cases, the class of
triangulations employed in dynamical triangulations is not only
triangulations of manifolds, but is rather ``psuedo-manifolds",
triangulations made from gluings \cite{Ambjorn:1996ny}. These may
not be simplicial complexes. However, distance along the skeleton
is still defined, as is inclusion of sub-simplices in higher
dimensional simplices, and so the process about to be described is
defined for these gluings.} $C_f$.  The set of vertices of $C_f$ will be
referred to as $V_f$, and the number of them as $N_f$. The
procedure starts with the random selection of a subset $V_c \subset V_f $ made up of $N_c$  vertices, which are to
serve as the vertices for the coarse-grained complex $\cc(V_c,C_f)$. This amounts to restricting the coarse-grained vertices
to lie on the vertices of the fine-grained triangulation.  In
order to approximate the uniform measure on sets of points in the
continuum, the probability of selecting a particular vertex is
weighted by its ``share of volume'', which in dynamical
triangulations is proportional to the number of highest
dimensional simplices containing the vertex (in 2D this is equal
to the valency of the vertex).  As above, the coarse-grained
complex can be written as a random variable $\cc(N_c,C_f)$, which
is abbreviated to $C_c(C_f)$ when a fixed $N_c$ is assumed.  The
``coarse-graining ratio'' $N_f/N_c$ will also be used below.

Similarly to the continuum case, we now associate a ``Voronoi
neighbourhood'' $\om(v_i)$ to each coarse-grained vertex $v_c \in V_c$.
The neighbourhood $\om(v_c)$ is in this case a subset of the
fine-grained vertices $V_f$. Here, $\om(v_c)$ contains
the set of fine-grained vertices that are closer to $v_c$ than
any other member of $V_c$, in graph distance on the
skeleton of $C_f$.

\begin{figure}[ht]
\centering
\resizebox{4.25in}{4.75in}{\includegraphics{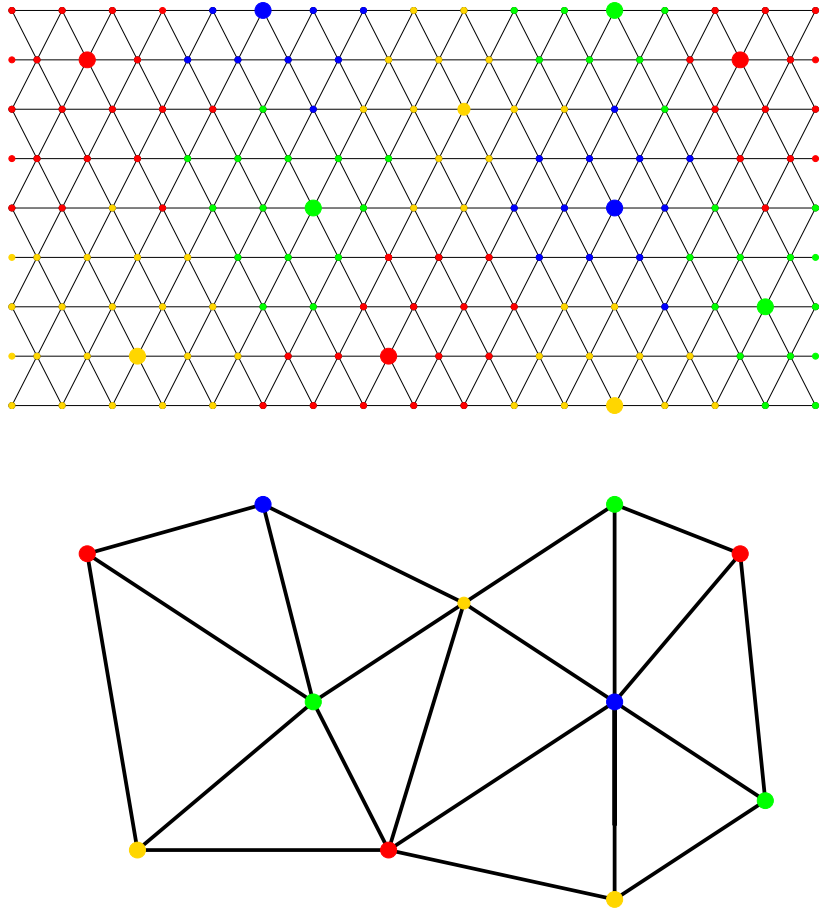}}
\caption{\small{ An instance of the coarse-graining procedure on a
triangulation. Here the fine-grained complex is a regular
triangular lattice. The vertices are shown as coloured dots, with
thin lines as edges. The coarse-grained vertices are shown as the
larger dots, and their Voronoi neighbourhoods are indicated by
colour. Below the main diagram, the coarse-grained vertices are
shown again, with the edges of the coarse-grained complex as think
black lines.  Note that some of the fine-grained points are
equidistant from many coarse-grained vertices, and had their
Voronoi neighbourhoods assigned randomly.  However, in this case,
the placement of edges in the coarse-grained graph was unaffected
by this.  A smaller proportion of fine-grained vertices are
randomly assigned when the ratio $N_f/N_c$ is
larger.}\label{f:grid1}}
\end{figure}

However, some fine-grained vertices can be equidistant from many
members of $V_c$ with non-zero probability.  A random
attribution of Voronoi neighbourhood is used for these vertices.
Consider a case where this occurs for a fine-grained vertex
$v_f \in V_f$, with equal distance $d$ from some set of nearest
coarse-grained vertices. There is a set of vertices that are
distance $d-1$ from a coarse-grained vertex, and which are
connected to $v_f$ by edges.  Call this set of vertices
$n(v_f) \subset V_f$. One member, $v'_f$, of $n(v_f)$ is selected at
random (uniformly), and $v_f$ is assigned to the same Voronoi
neighbourhood as $v'_f$. This assumes that the neighbourhood of
$v'_f$ has already been decided, and so the process must be
followed for vertices at the smallest distance $d=1$, outwards.
This is how the problem is dealt with in general. Starting from
those nearest to a coarse-grained vertex, all fine-grained
vertices can be assigned a Voronoi neighbourhood in this way\footnote{This method is chosen in order to increase the number of cases in which a simplicial manifold is formed by the coarse-grained complex.  For instance, if the
fine-grained history was a regular flat triangulation, we would
want the coarse-grained version to be a 2 dimensional
triangulation with high probability.  This is the reason why $v_f$ is randomly assigned to one of the 4 possible neighbourhoods, rather than adding more edges. This will more often produce a 2D triangulation in these circumstances. See figure \ref{f:grid1} for an example.}.

Now the edges of $\cc(C_f)$ are added. In the continuum procedure,
the edges are between coarse-grained vertices whose Voronoi
neighbourhoods share boundary points.  In this graph procedure,
the different neighbourhoods are connected by edges, which play
the role of the boundaries. Analogously to the continuum case, an
edge is added in the coarse-grained graph between $v_c$ and
$w_c$ iff there is an edge in $C_f$ between any member of
$\om(v_c)$ and any member of $\om(w_c)$. This defines
the skeleton of $\cc(N_c,C_f)$.

The addition of higher order simplices is similar.  In
$\cc(N_c,C_f)$, there is an $n$-simplex comprised of the coarse-grained vertices in set $b \subset V_c$ if their neighbourhoods ``share an n-simplex'', \ie if there is an $n$-simplex $s$ in $C_f$ such that

\begin{equation}
s  \cap \om(v_c) \neq \emptyset \quad \forall \, v_c \in b.
\end{equation}

The process is illustrated in figure \ref{f:grid1}.

It is also easy to define part of a variant to the scheme: a
similar process on the cell complex dual to the triangulation. The
definition of the skeleton of the coarse-graining would be
entirely analogous, and that is all that is needed for some of the
interesting observables, like the proposed effective average
scalar curvature. It would make a good consistency check to
develop two coarse-graining schemes which could be compared, and
the dual version may be more practical in cases where
configurations in the CDT simulations are stored as the dual
complex (this is the case in the 3D simulations
\cite{Ambjorn:2000dj}).

Being applicable to any simplicial complex, this would be possible to apply to dynamical triangulations. So far the discussion has been appropriate to Euclidean dynamical triangulations, but the procedure could also be applied to Wick
rotated causal dynamical triangulations. After coarse-graining, the special layered structure of the CDT is lost, but that has no bearing on the recovery of physically interesting observables. A further difference in the case of causal dynamical triangulations is that there are, in the 4D model, two possible edge lengths rather than just one, for spacelike and timelike edges.  This complication is ignored for now (as it has been when calculating the spectral dimension) as it unlikely to grossly affect properties of interest. But in principle it would not be hard to incorporate different edge lengths into the coarse-graining procedure.

\subsection{The application to a CDT path integral}

\label{s::sketch}

It is helpful to have a definite scheme laid out for the application to CDT simulations, which is is next step for this program.

The methods of computing the coarse-grained observables will be very similar to those used to find the spectral dimension and other observables.  The discretised CDT path integral at fixed volume has been given in equation (\ref{e::CDTpi}).  Monte-Carlo simulations produce instances of causal dynamical triangulations according to the weights given by the Euclideanised path integral, which can be used to estimate the expectation values of observables at different values of the cut-off.  The approach is be similar to that used in lattice QFT.  In the CDT simulations, the number of coarse-grained vertices $N_c$, which is proportional to the dimensionful volume, will be held fixed while the number of fine-grained simplices is moved towards infinity.  The remaining dimensionless coupling constant of the model is the inverse (bare) Newton's constant which will be held fixed.  As the lattice spacing is sent to zero, large-scale effective observables should approach a fixed value, their continuum limit. Using the techniques of the previous section, certain Delaunay complex observables of section \ref{s::observables} will be estimated.

It would be sensible to start by applying the idea to the 3D model.  In 3D, the only relevant solution of the Einstein equations is De Sitter, or on the Euclidean side the sphere\footnote{Because of the topological restrictions the simulations
cannot strictly produce a sphere, but it is argued that a sphere
connected at two points by a cylinder of minimal diameter should
play the same role as the classical solution \cite{Ambjorn:2000dja}.}. So far all measured observables are consistent with the conjecture that the 3D
simulations are producing spheres with small fluctuations in the
geometry.  Because of the lack of propagating degrees of freedom in 3D gravity, it is hoped that no spatially correlated geometrical fluctuations will survive in the continuum limit, \ie{} that the limiting geometry will be exactly a sphere. Applying the coarse-graining scheme, the first aim would be to compare the expectation values of certain Delaunay complex observables (such as the average valency) from the CDT simulations with the values obtained from a sphere with the same number of coarse-grained vertices.  In the limit, the hope would be that the CDT expectation values converge to the values for the sphere.  Fortunately, Delaunay triangulations of a sphere are easily generated by computer, making Delaunay complex observables easy to compute in this case, even if they cannot be found analytically.

The observables mentioned in section \ref{s::examples} could be used for this purpose.  Beyond this, it would be desirable to know a \textit{sufficient} list of Delaunay
observables $\{O_i\}$, such that the sphere (and geometries
indistinguishable from it at large scales) is picked out as the
only geometry approximately holding a certain set of values
$O_i=O_i^*$.  In 3D perhaps this could be explored analytically, building on existing results \cite{isokawa:2000}. 

If successful, similar techniques would then be applied in 4D.  Also, the new scaling dimensions and the measures of effective manifoldlikeness discussed in section \ref{s::observables} would be interesting to calculate for 4D CDT's, and the results compared to the other dimension estimators.

\section{Conclusion}

A new method for coarse-graining dynamical triangulations has been
presented, and some arguments for its physical suitability have
been given.  As in other aspects of the theory of dynamical
triangulations, the aim has been to build on ideas from
standard lattice quantum field theory. Thus, as in
lattice quantum field theory, coarse-grained observables can be
calculated for each history in the quantum sum, and the
expectation value is found as a statistical average. In the
lattice field theory case, it is crucial that the coarse-grained
version of an observable only depends on large scale properties of
the history, and this remains crucial for quantum gravity
coarse-graining schemes, where it becomes a more involved issue. This
must be a primary consideration for any such scheme.

A number of technical issues remain. There is a need for extensions of known results of \cite{Bombelli:2004si,isokawa:2000} that relate average valency of the Random Delaunay complexes to scalar curvature.  Their conjecture is that all scalar curvature invariants can be calculated from simple
properties of random Delaunay complexes in 3 and 4 dimensions (it
is proved in 2), for manifolds of low curvature on the sprinkling
scale.  It will be useful for this coarse-graining program to
find these expressions for the curvature invariants.  Other
conjectures are raised in this paper to add to those of Bombelli,
Corichi and Winkler. It would be interesting to extend their results to
geometries that are ``close'' to the low curvature
geometries that have so far been studied.  This would amount to a
justification of the central conjecture given in section
\ref{s::observables} for these observables.  Finding a characterisation of spheres in terms of simple properties their random Delaunay triangulations would be useful for the 3D and 4D simulations.

It would also be of use to carry out some computational tests of
the scheme set out above. Some of the above conjectures are amenable to this approach.  For example, a 2D torus with extended dimensions, and a 3D torus with one compactified dimension, are uncontroversially ``close'' geometries in the relevant sense.  Tests could be performed to see if interesting Delaunay observables converged to the same values for the these two types of geometries at small sprinkling density.  This could also be done for the discrete coarse-graining of section \ref{s::practical}.

Adding matter to the scheme would be fairly easy in principle. For
instance, for Ising matter the spin of a coarse-grained vertex $v_c$ could be taken
directly from the value on the fine-grained history (similar to
Kadanoff's decimation method \cite{Wilson:1974mb}) .  Alternatively, the mode
spin of vertices in the discrete Voronoi neighbourhood
$\om(v_c)$ defined in section \ref{s::coarseg} could be taken as
the spin for the coarse-grained vertex $v_c$  (similar to the
majority spin method).

The coarse-graining procedure may also have uses outside of
dynamical triangulations, as it can be applied to any simplicial
complex (indeed, this might even suggest new ways to discretise the path integral). Unlabelled graphs were considered as encoding geometry in
\cite{Bombelli:2004si}, as a stepping stone to the spin-networks
of loop quantum gravity.  It is possible that this scheme would be
of use in this context as well, and that a similar scheme could be
developed for labelled graphs.  In this case however, the issue of
non-local links \cite{Markopoulou:2007ha} may need to be dealt with. It would be interesting to see if some modification of the scheme was eventually
necessary.  The application of this, or similar coarse-graining ideas, in the
spin-foam program (considered as a path-integral formulation of QG as in \cite{Oriti:2001qu})  is also a possibility.

This paper is intended as the first in a series.  The next step is the application of the coarse-graining procedure, as outlined in section \ref{s::sketch}.  The final worth of any coarse-graining scheme depends on the success of the calculations in producing good, consistent results.  On the point of comparison with results, further adaptation of the scheme may become necessary, similarly to other successful lattice coarse-graining programs, as mentioned in section \ref{s::geometry}.  From simulations so far, indications are that the program should be practical, in terms of available computational power.  In this way, hopefully, evidence for the convergence of renormalised physical quantities like average scalar curvature could be found.

\section*{Acknowledgements}

The author would like to thank Jan Ambj\o rn, Renate Loll, Sumati Surya and Pedro Machado for helpful comments.  Some simulations that informed this work used the QHULL package.  The work was carried out in part at the Institute for Theoretical Physical at the University of Utrecht.  Research at Perimeter Institute for Theoretical Physics is supported in part by the Government of Canada through NSERC and by the Province of Ontario through MRI.

\bibliographystyle{h-physrev3}
\bibliography{refs}

\appendix

\section{Observables for Manifoldlikeness in 2D}

\label{a::2Dtriangulations}

It is claimed in section \ref{s::examples} that, if a geometry
$X$ is effectively manifoldlike at a certain scale, then a random
Delaunay complex on $X$, taken at the appropriate sprinkling
density, will be a triangulation of a manifold.  Effective
manifoldlikeness is a very interesting observable for quantum
gravity.

However, this easily stated condition of manifoldlikeness has no
tolerance. The addition or subtraction of one edge can alter the
result. In simulations, small deviations are bound to arise at
some scales, and so we will need to quantify how close a graph is
to being a triangulation. One such quantification for a graph $G$
is the size of the largest induced sub-complex in $G$ that is also
an induced sub-complex of some triangulation.  Another quantity
would be the number of vertices whose ``star'' (defied below) is
topologically a ball of given dimension.  Below, similar
quantifications are introduced that are easy to calculate.

At a practical level, it is preferable to deal only with the
skeleton of the complex where possible, to avoid having to compute
the higher dimensional simplices.  Also, it is possible that
measures of manifoldlikeness depending only on the skeleton would
be more robust against small deviations of the type discussed
above.  With this in mind, we would like to distinguish graphs
which are the skeleton graph of some triangulation of a manifold.
These questions are considered in this appendix for the simple
case of 2D triangulations.

First, it is necessary to be more specific about the kinds of
triangulation under consideration.  An abstract simplicial complex
has been defined in section \ref{s::delaunay}, and we continue to
drop the term ``abstract'' and treat it as implicit in the
following.  A 1D simplicial complex is a graph $G=\{S_0,S_1\}$,
where $S_0$ is the set of vertices and $S_1$ is the set of edges.
In 2D, we have $S=\{S_0,S_1,S_2\}$ where $S_0$, $S_1$, and $S_2$,
are the vertices, edges and triangles. Where definitions below
differ from those in \cite{Ambjorn:1996ny}, it is a consequence of
the use of the abstract description of simplicial complexes.

A triangulation of a manifold is called a simplicial manifold,
which is a type of simplicial complex.  To characterise them
further we need the following definitions.  In a simplicial
complex, the \textit{star} of a simplex $\sigma$, $\st(\sigma)$,
is the the set of all simplices of which $\sigma$ is a face, and
their faces.  The star is also a simplicial complex by this
definition. The link of $\sigma$, $\li(\sigma)$, is the set of all
simplices in the star of $\sigma$ for which $\sigma \cap \sigma_f
= \emptyset$. The following is an immediate consequence theorem 1
of \cite{Ambjorn:1996ny}.

\begin{theorem}

\label{t::frombook}

a 2D simplicial complex $S$ is a simplicial manifold iff the link
of every vertex in $S$ is a cycle of edges.

\end{theorem}

To rephrase the original problem, we want to be able to
distinguish those graphs $G=\{S_0,S_1\}$ for which there exists a
simplicial manifold $S(G)=\{S_0,S_1,S_2\}$. The task will involve
identifying $S_2$, an assignment of triangles to the graph, if one
exists.

First, we must examine what theorem \ref{t::frombook} means for
the skeleton of the simplicial manifold.

\begin{figure}[ht]
\centering
\resizebox{3.0in}{1.0in}{\includegraphics{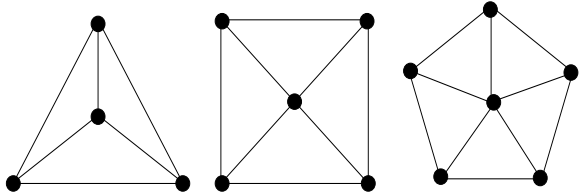}}
\caption{\small{ Wheel graphs, of order 4, 5, and
6.}\label{f::wheels}}
\end{figure}

A \textit{wheel} is a graph of order $n$ which contains a cycle of
order $n-1$, and for which every graph vertex in the cycle is
connected to one other graph vertex, which is known as the hub
(see figure \ref{f::wheels}). A \textit{full wheel} of a graph $G$ is a subgraph
of $G$ that is a wheel and contains all the vertices of $G$; a graph $G$ will be said to have a full wheel if it contains a full wheel of itself.  A few easily proved lemmas are necessary for the final results.

\begin{lemma}

\label{l::wheel}

For any vertex $v$ in a 2D simplicial manifold, the skeleton of
$\st(v)$ is a wheel of which $v$ is the hub.

\end{lemma}

\proof  Consider $v$, a vertex in a 2D simplicial manifold. Clearly, the vertices in $\st(v)$
are exactly those in $\li(v)$ and $v$.  The skeleton of $\li(v)$
is a cycle, from theorem \ref{t::frombook}, and so the skeleton of
$\st(v)$ contains a cycle containing all vertices but $v$. Every
vertex in the cycle $\li(v)$ shares an edge with $v$, so the
skeleton of $\st(v)$ is a wheel with $v$ as the hub. $\Box$

\paragraph{} Now we have characterised the skeleton of $\st(v)$ for $v$ a
vertex in a simplicial manifold.  But the definition of $\st(v)$
depends on more than the skeleton $G$ of the simplicial manifold.
To use lemma \ref{l::wheel} we must be able to identify the
skeleton of $\st(v)$ in $G$ without reference to any triangles.  The concept of the \textit{graph neighbourhood} is useful here. Let the graph neighbourhood,
$\gn(v)$, of a vertex $v$ in a graph $G$ be the subgraph induced
by $v$ and all of those vertices that share an edge with $v$.

\begin{lemma}

\label{l::wheelg}

For any vertex $v$ in a 2D simplicial manifold, $\gn(v)$ has a full wheel, of which $v$ is the hub.

\end{lemma}

\proof Vertices in $star(v)$ either share an edge with $v$, or are
$v$.  Conversely all vertices that share an edge with $v$ are in
$\st(v)$.  Therefore the skeleton of $star(v)$ is contained in
$gn(v)$ and $gn(v)$ contains no other vertices. The lemma then
follows from lemma \ref{l::wheel}. $\Box$

\begin{figure}[ht]
\centering \resizebox{1.3in}{1in}{\includegraphics{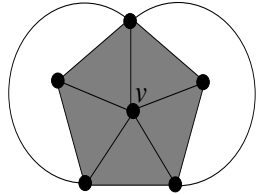}}
\caption{\small{ An example of a graph containing a vertex $v$,
that has a unique full wheel subgraph of which $v$ is the hub. The set
of triangles $t(v)$ is shown: the triangles in $t(v)$ are the
boundaries of the shaded regions. This $t(v)$ is the set of
3-cycles containing $v$ in the full wheel subgraph. Note that not
all 3-cycles in the graph are in $t(v)$, as they are not all in
this wheel. By assigning these triangles to all graph
neighbourhoods in a graph, it can be seen if the graph is the
skeleton of a simplicial manifold.}\label{f::graphn}}
\end{figure}

\paragraph{} On its own, this provides a necessary condition for a graph $G$ to
be the skeleton of a simplicial manifold.  It is also possible to find a sufficient condition.
Let us take a graph $G$ such that the condition of lemma \ref{l::wheelg} is satisfied, and such that there is one and only one full wheel in $gn(v)$ for all $v$. Each vertex in
$G$ then has a unique associated wheel graph. We may then
associate a unique set of triangles $t(v)$ to each vertex: the set
of 3-cycles in this wheel.  This is shown in figure \ref{f::graphn}.

\begin{theorem}

\label{t::graphtosm}

Let $G=\{S_0,S_1\}$ be a graph such that, for every $v$, $\gn(v)$
has one and only one full wheel subgraph of which $v$ is the hub.  Let $t(v)$ be the set
of 3-cycles in this wheel, and let $S(G)=\{S_0,S_1,S_2\}$ be the simplicial complex such that $S_2$ is the union of $t(v)$ over all $v \in G$.  If $t(v)$ are the only triangles containing $v$ for all $v$, then $S(G)$ is a 2D
simplicial manifold.

\end{theorem}

\proof Consider a vertex $v$ in a simplicial complex $S(G)$ as
described in the theorem. If $t(v)$ are the only triangles
containing $v$ in $S(G)$, then they are the only triangles in
$\st(v)$ in $S(G)$.  It is easy to see that $\li(v)$ is therefore
a cycle.  By theorem \ref{t::frombook}, if this is true for all
vertices in $S(G)$, then $S(G)$ is a 2D simplicial manifold.
$\Box$

\paragraph{} Now we have a sufficient condition for a graph $G$ to be the
skeleton of some simplicial manifold. In a simulation, therefore,
only a positive result would be conclusive. To make a necessary
condition, we would have to allow $\gn(v)$ to contain more than
one full wheel subgraph, and look for an assignment of $t(v)$ such
that $S(G)$ is a 2D simplicial manifold.  But we will now show
that the class of simplicial manifolds which do not satisfy the
condition of theorem \ref{t::graphtosm} is not important for our
purposes.

\begin{lemma}

\label{l::graphn}

The graph neighbourhood $\gn(v)$ of every vertex $v$ in the
triangulation of a sphere is a planar graph, which has a full
wheel subgraph of which $v$ is the hub.

\end{lemma}

\proof It is well known and easy to see that the skeleton $G$ of a
triangulation of a sphere is a planar graph in which every region
is bounded by three edges. The graph neighbourhood of a vertex in
$G$ is a subgraph of $G$, and is therefore also planar. The lemma
then follows from lemma \ref{l::wheelg} . $\Box$

\begin{lemma}

\label{l::onewheel}

If $\gn(v)$ in a graph $G$ is planar and has a full wheel subgraph
of which $v$ is the hub, then it has only one full wheel subgraph
of which $v$ is the hub.

\end{lemma}

\proof In a graph $G$, consider a graph neighbourhood $\gn(v)$ of
a vertex $v$ that is planar and has a full wheel subgraph, of
which vertex $v$ is the hub.  The lemma is trivial when $\gn(v)$
contains 4 vertices, as the full wheel subgraph is then the
complete graph on 4 vertices.

Consider the case in which $\gn(v)$ is of order 5.  Let us label
the vertices $\{v,1,2,3,4\}$.  The graph $\gn(v)$ contains edges
from $v$ to all other vertices.  It also contains a cycle from the
full wheel graph which we can label
$\{\{1,2\},\{2,3\},\{3,4\},\{4,1\}\}$ or schematically
$1-2-3-4-1$. In order for $\gn(v)$ to contain \textit{more} than
one full wheel with $v$ as hub, it must also contain another
4-vertex cycle in the subset $\{1,2,3,4\}$, and therefore would
contain the edges $\{2,4\}$ and $\{1,3\}$.  In this case, $\gn(v)$
is the complete graph on 5 vertices, $K_5$. This is not a planar
graph, and so this violates our assumptions.

For larger order $\gn(v)$, it is not hard to see that any other
$\gn(v)$ with two full wheel subsets (with the same hub) can be
reached from $K_5$ by graph expansion.  Similarly to the order 5
case, let us label the vertices $\{v,1,2,3,4,...,n\}$ and let one
of the cycles $c$ be $1-2-3-4-...-n-1$.  There is another order
$n$ cycle in $\gn(v)$, $\tilde{c}$ containing $\{1,2,3,4,...,n\}$,
$p_1-p_2-p_3-...-p_n-p_1$, where $\{p_i\}$ is a non-cyclic
permutation of $\{1,2,3,4,...,n\}$.  From the properties of
permutations, there exists some pair of edges $\{p_a,p_b\}$ and
$\{p_c,p_d\}$ in $\tilde{c}$ such that $p_a<p_c<p_b<p_d$ or
$p_d<p_a<p_c<p_b$. Therefore, all pairs of vertices in the set
$\{v,p_a,p_b,p_c,p_d\}$ are connected by paths that do not include
any other vertex in the set $\{v,p_a,p_b,p_c,p_d\}$ (this is
because $v$ shares an edge with all others, $\{p_a,p_b\}$ and
$\{p_c,p_d\}$ are edges, and the rest are connected by the cycle
$c$). It follows from this that $K_5$ is a graph minor of $\gn(v)$
(\ie{} related to $\gn(v)$ by edge deletion and/or edge
contraction). Therefore such $\gn(v)$ are not planar by
Kuratowski's reduction theorem, and violate our assumptions for
the same reason. $\Box$

\begin{corollary}

The graph neighbourhood $\gn(v)$ of every vertex $v$ in the
triangulation of a sphere has has one and only one full wheel
subgraph of which $v$ is the hub.

\end{corollary}

This corollary is the main result of this appendix.  It shows that the graph neighbourhoods in the triangulation of a sphere satisfy the conditions in theorem
\ref{t::graphtosm}. This condition is therefore sufficient for a graph $G$ to be the skeleton of a simplicial manifold, and necessary for it to be the skeleton of a triangulation of a sphere.

If a graph neighbourhood $\gn(v)$ in a graph
$G$ does contain more than one full wheel graph, then it is not
isomorphic to a graph neighbourhood in any triangulation of a
sphere; $G$ contains a ``small handle'' at $v$.  If this $G$ arose
from the random Delaunay process on a manifold $X$, this can be
counted as short scale detail in $X$, and so it is justified to
say that $X$ is not manifoldlike at that scale -- this topological
detail should be removed by decreasing the sprinkling density.
Also, spacetime topology is fixed in the CDT models that are presently being
used in simulations, and so we do not expect that such handles will arise
there. With this in mind, we expect that the condition of theorem
\ref{t::graphtosm} is not too strict to be of use in CDT
simulations.

This condition also suggests two measures of how close a graph is
to being a triangulation.  The first, $\TA(G)$, is the fraction of
vertices $v$ such that $\gn(v)$ is planar and contains a full
wheel of which $v$ is the hub.  The second, $\TB(G)$, is the
fraction of vertices $v$ in $G$ that obey this condition, plus the
extra condition that the $t(v)$ are the only triangles containing
$v$. Only if $\TB(G)=1$ do these measures show that $G$ is the
skeleton of a simplicial manifold, by theorem \ref{t::graphtosm}.
For a dynamical triangulations simulation, if the average value of
$\TB$ approached 1 as the lattice spacing $a$ was taken to zero,
this would be evidence of manifoldlike behaviour on the sprinkling
scale, according to the conjectures of section
\ref{s::observables}.  We can add to this list the more strict
measure $\TC(G)$ which is 1 if $G$ satisfies the conditions of
theorem \ref{t::graphtosm} and 0 otherwise.

These quantities have been found to be practically computable for large simplicial complexes generated by the Delaunay procedure on a 2D sphere.  For higher
dimension, the conditions will inevitably be more complicated,
but may still be practical to use.  In this case, a triangulation
of a sphere is not uniquely determined by its skeleton.  The
alternative, of computing the link given the higher dimensional
simplices, may then be preferable.  This choice will be affected
by the confidence in different aspects of the coarse-graining
scheme, and this can only be decided as the scheme is further examined.

\end{document}